\documentclass[%
 reprint,
superscriptaddress,
 amsmath,amssymb,
 aps,%prl
floatfix,
]{revtex4-1}

\usepackage{graphicx}% Include figure files
\usepackage{dcolumn}% Align table columns on decimal point
\usepackage{bm}% bold math
\usepackage{soul,color}
\usepackage{float}
\usepackage{multirow}
\begin{document}

\preprint{APS/123-QED}

%\title{On-chip microwatts-threshold optical parametric oscillator \\ using periodically poled lithium niobate microring resonators}% Force line breaks with \\
%\title{On-chip lithium niobate optical parametric oscillator with micro-watts threshold }% 

\title{Ultralow-threshold thin-film lithium niobate optical parametric oscillator}% 

\author{Juanjuan Lu}
\author{Ayed Al Sayem}
\author{Zheng Gong}
\author{Joshua B. Surya}
\affiliation{Department of Electrical Engineering, Yale University, New Haven, Connecticut 06511, USA}
\author{Chang-Ling Zou}
\affiliation{Department of Optics, University of Science and Technology of China, Hefei 230026, P. R. China.}
\author{Hong X. Tang}
 \email{hong.tang@yale.edu}
\affiliation{Department of Electrical Engineering, Yale University, New Haven, Connecticut 06511, USA}

\date{\today}% It is always \today, today,

\begin{abstract}
 Materials with strong $\chi^{(2)}$ optical nonlinearity, especially lithium niobate, play a critical role in building optical parametric oscillators (OPOs). However, chip-scale integration of low-loss $\chi^{(2)}$ materials remains challenging and limits the threshold power of on-chip $\chi^{(2)}$ OPO. Here we report the first on-chip lithium niobate optical parametric oscillator at the telecom wavelengths using a quasi-phase matched, high-quality microring resonator, whose threshold power ($\sim$30\,$\mu$W) is 400 times lower than that in previous $\chi^{(2)}$ integrated photonics platforms. An on-chip power conversion efficiency of 11\% is obtained at a pump power of 93\,$\mu$W. The OPO wavelength tuning is achieved by varying the pump frequency and chip temperature. With the lowest power threshold among all on-chip OPOs demonstrated so far, as well as advantages including high conversion efficiency, flexibility in quasi-phase matching and device scalability, the thin-film lithium niobate OPO opens new opportunities for chip-based tunable classical and quantum light sources and provides an potential platform for realizing photonic neural networks.
\end{abstract}

\maketitle

%\tableofcontents
Optical parametric oscillators (OPOs) are critical sources of coherent radiation over a wide spectral range and have been workhorses for spectroscopy in the near- and mid-infrared regime~\cite{dunn1999parametric,https://doi.org/10.1002/lpor.201100036,Maidment:16}. They are also widely used as non-classic light sources for applications in quantum information processing, including quantum random number generation, quantum key distribution, and recently demonstrated Ising machines~\cite{Andersen_2016,madsen2012continuous_QKD,inagaki2016large,isingmachine}. Microcavity-based OPO has been particularly attractive due to its miniaturization and power-efficiency. With ultra-high optical quality (Q) factor of $10^7$–$10^8$ afforded by the whispering gallery resonator (WGR) system, low-threshold OPOs have been achieved in various bulk WGRs~\cite{PhysRevLett.93.083904,PhysRevLett.93.243905,LNWGMOPO,PhysRevLett.106.143903,Herr:18,sayson2019octave,AgGaSe2_opo}, especially the highly polished lithium niobate (LN) WGRs with $\chi^{(2)}$ nonlinearity.  

Over the past decade, the demand for scalability in photonic components has stimulated efforts to produce on-chip OPOs via either $\chi^{(2)}$ or $\chi^{(3)}$ nonlinearity~\cite{XiyuanLu_OPO,Lu:20,ji2017ultra,SiC_opo,chang2020ultra,Bruch_OPO}. Among these, $\chi^{(2)}$ OPO has only been demonstrated with an aluminum nitride (AlN) microring resonator while its threshold power is tens of milliwatts~\cite{Bruch_OPO}, which is orders of magnitude larger than state-of-the-art value of lithium niobate WGR. Hence, low threshold and scalable $\chi^{(2)}$ OPO remains challenging. Recently, thin film LN has emerged as a compelling integrated nonlinear and quantum photonics platform, whose favorable performances including but not limited to electro-optic modulation~\cite{loncarmodulator,zhang2019broadband,phcmodulator}, frequency comb generation~\cite{wang2019monolithic,okawachi2020chip,Gong:20}, frequency conversion~\cite{Wang2018g,Chen:19,Lu2019d,lu20201}, and entangled photon pair generation~\cite{PhysRevLett.124.163603,ma2020ultrabright}, have by now been well demonstrated. 

In this Letter, we present an ultralow-threshold, chip-integrated OPO using a periodically poled lithium niobate microring resonator (PPLNMR). With high optical confinement and strong spatial mode overlap via the quasi-phase matching, a threshold power as low as 30\,$\mu$W is demonstrated for the parametric oscillation at the infrared band, which is a state-of-the-art value reported among all current integrated photonics platforms. Meanwhile, an absolute power conversion efficiency of 11\% is obtained at a pump power of 93\,$\mu$W. A tuning bandwidth of 180\,nm (23\,THz) could be addressed through a small change of 5\,nm (2.5\,THz) in pump wavelength (frequency). Alternatively, the temperature tuning is investigated to fine control the signal and idler wavelengths and a degenerate 
OPO is achieved at an optimal temperature. Overall, low-threshold, wide-tunability, as well as flexibility in phase-matching by domain poling make thin-film LN OPO the ideal candidate for compact, versatile light sources for various on-chip optical applications, such as sensing, frequency comb spectroscopy, and quantum squeezers.

\begin{figure*}[htbp]
\includegraphics[width=0.9\textwidth]{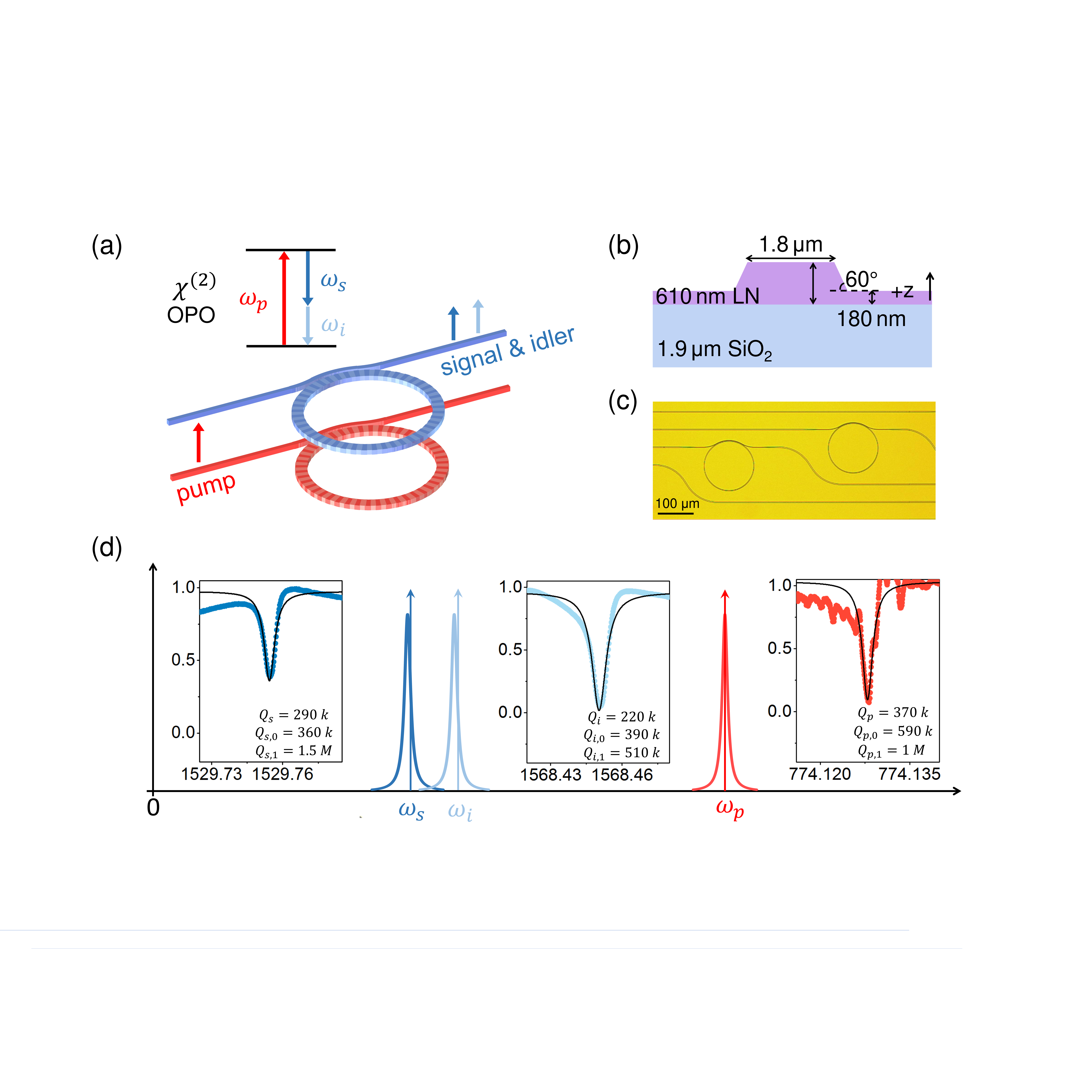}  
\caption{\label{intro}\textbf{Design and fabrication of a quasi-phase-matched optical parametric oscillator}. (a) Illustration of the parametric oscillation using a $\chi^{(2)}$ PPLNMR. Signal and idler lights are generated by a near-visible pump. $\omega_p$, $\omega_s$ and $\omega_i$ represent frequencies for pump, signal and idler light, respectively. (b) Cross-sectional view of the microring indicating the key geometric parameters. (c) Optical image of the fabricated PPLNMR devices. (d) Schematics of the mode frequencies and their respective transmission spectrum with the extracted Q values shown in the inset. The subscripts $0,1$ represent the intrinsic and coupling Q factors, respectively.}
\end{figure*}
Figure\,\ref{intro}(a) illustrates the principle of our lithium niobate microring OPO device, in which $\chi^{(2)}$-based parametric down-conversion process allows the generation of the infrared signal and idler lights with a near-visible pump. $\omega_p$/$\omega_s$/$\omega_i$ represent the pump/signal/idler mode frequencies with their energy conservation ($\omega_p=\omega_s+\omega_i$) shown in the inset. When the phase-matching and triply-resonant conditions are satisfied for these three optical modes, the parametric process would be enhanced by the high-Q resonator. The theoretical pump power threshold $P_{th}$ for parametric oscillation could be obtained when the optical loss and parametric gain are balanced~\cite{Bruch_OPO}:
\begin{equation}
\label{eq_threshold}
P_{th}=\frac{\hbar \omega_p}{8g^2}\frac{\kappa_s \kappa_i \kappa_p^2}{\kappa_{p,1}}=\frac{\hbar \omega_s \omega_i \omega_p^2}{64 g^2}\frac{Q_{p,1}}{Q_s Q_i Q_p^2},
\end{equation}
where $g$ is the vacuum coupling strength between the three interacting modes, $Q(\kappa)_{s/i/p}$ is their respective loaded Q factor (total dissipation rate), and $Q(\kappa)_{p,1}$ is the coupling Q factor (external dissipation rate) of the pump mode. Therefore, a large $g$ and high Q factors are always demanding to reduce the OPO threshold power.

In the microring resonator, the coupling rate $g$ is dependent on the material $\chi^{(2)}$ coefficient and modal overlap factor $\gamma$ through the relation: $g\propto \chi^{(2)}\gamma $~\cite{Lu2019d}. Here, to optimize $g$, we utilize the largest $\chi^{(2)}$ tensor element $d_{33}$ of z-cut LN thin film, and also improve the mode overlap factor $\gamma$ close to unity by employing fundamental transverse-magnetic modes. In contrast to previous study of on-chip OPO in AlN platform~\cite{Bruch_OPO}, where high-order mode is leveraged to compensate the dispersion at the visible wavelength, we realize the quasi-phase matching (momentum conservation) between fundamental modes via the radial periodic-poling of the LN microring. Accordingly, for a microring whose key geometric parameters are indicated in Fig.\,\ref{intro}(b), an azimuthal grating number $M=146$ (poling period $\Lambda$=2$\pi R/M$=3\,$\mu$m) is designed to compensate for the momentum mismatch $\Delta m=|m_s+m_i-m_p|$ for the parametric oscillation at 1550\,nm with a pump at 775\,nm. The phase-matching of the device was verified through the reversal process of degenerate OPO, i.e. second-harmonic generation (SHG), as elaborated in our previous work~\cite{lu20201}. Figure\,\ref{intro}(c) is an optical image of the fabricated PPLNMR device (See Methods for fabrication details). The cavity transmission spectra of the signal, idler, and pump modes for the subsequent OPO are plotted in Fig.\,\ref{intro}(d), where Lorentzian fittings are applied to extract loaded Q factors of $2.9\times 10^5$, $2.2\times 10^5$, and $3.7 \times10^5$, respectively. Their intrinsic Q factors are calculated to be $3.6\times 10^5$, $3.9\times 10^5$, and $5.9\times 10^5$ at under-coupled conditions, which are confirmed by measuring the dependence of loaded quality factor on the coupling gap in an array of fabricated devices.

\begin{figure}[htbp]
\includegraphics[width=0.45\textwidth]{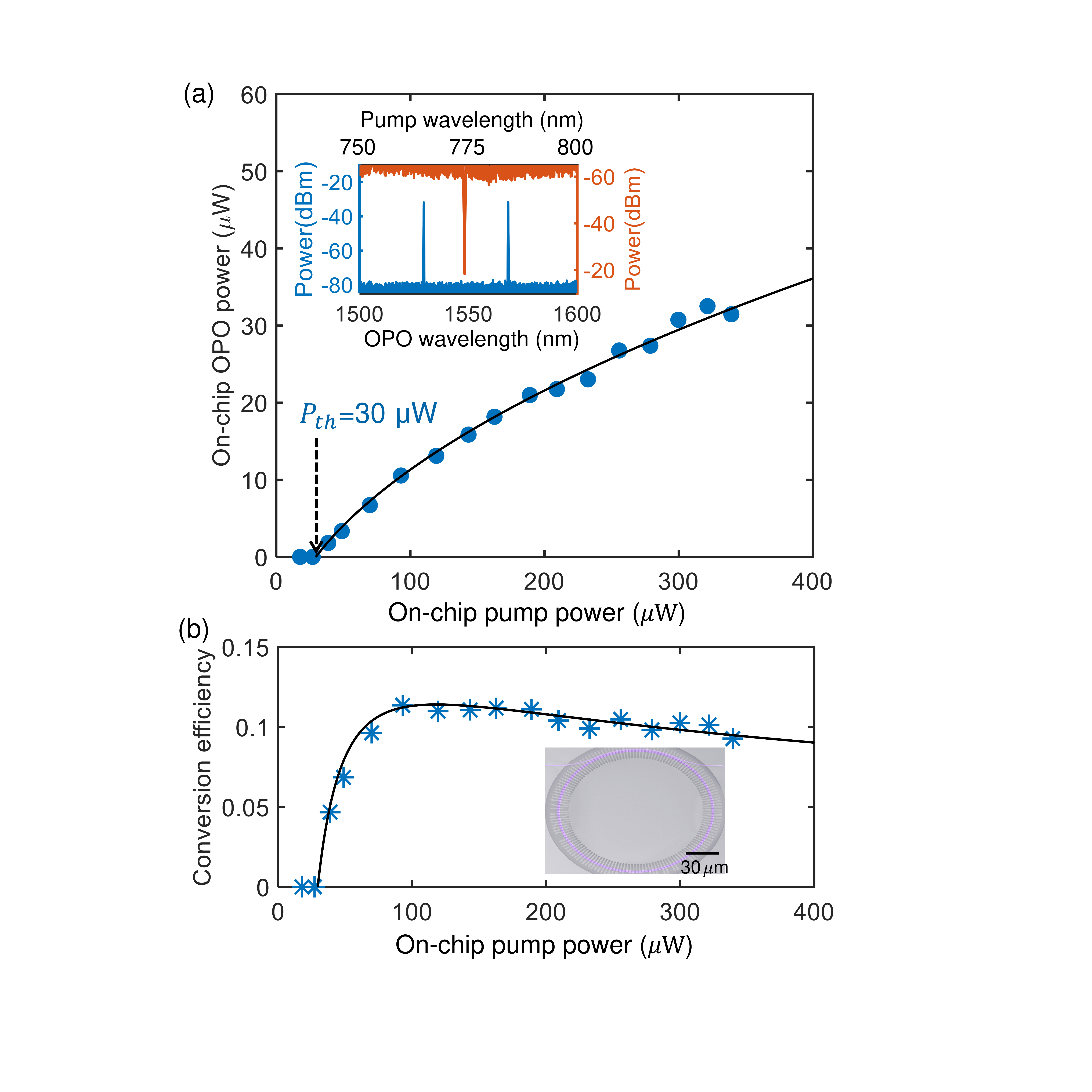}  
\caption{\label{opo} \textbf{Power dependence of the PPLNMR-based OPO}. (a) The experimental (blue circles) and theoretical (solid line) on-chip infrared power versus on-chip pump power for the parametric oscillation process. An example snapshot of off-chip pump (orange), and oscillation (blue) spectra after the device are shown in the inset. (b) On-chip conversion efficiency of pump to both
signal and idler fields versus pump power, where a theoretical fit (solid line) is applied. Inset: scanning electron micrograph of a PPLNMR mock-up etched in hydrofluoric acid.} 
\end{figure}

When pumping the device by a near-visible tunable laser (New Focus TLB-6712, 766--781\,nm), the parametric oscillation at the infrared wavelengths could be observed if the pump power exceeds the threshold. In practice, due to the resonance frequency mismatch ($\delta=\omega_p-\omega_s-\omega_i$) induced during the fabrication process, the OPO threshold rises to $P_{th}\cdot(1+\delta^2/\kappa_i\kappa_s)$. By controlling the chip temperature to adjust $\delta$ and varying the pump power, the parametric oscillation with the lowest threshold is observed when the device is pumped at the resonance near 774.1\,nm at an optimal temperature of 125$^\circ$C, as indicated in Fig.\,\ref{opo}(a). The measured on-chip OPO power (blue circles) is plotted against the on-chip pump power, where a threshold power $P_{th}$ of $\sim$30\,$\mu$W is identified and corresponds to an intra-cavity pump photon number of $\sim$5.4$\times 10^4$. The parametric oscillation spectrum at a pump power of 120\,$\mu$W is captured using an optical spectrum analyzer (Yokogawa AQ6374, 350--1750\,nm), where distinct signal and idler waves are resolved at 1529.7 and 1568.4\,nm respectively, as shown in the inset.

Substituting the measured threshold, and Q factors [Fig.\,\ref{intro}(d)] in Eq.\,(\ref{eq_threshold}), we extract a $g/2\pi$ of 1.2\,MHz, consistent with the value obtained via the independent SHG measurement~\cite{lu20201}. Theoretically, the power of the signal and idler OPO emission under the triply-resonant condition could be derived as \cite{Bruch_OPO}
\begin{equation}
\label{eq_p_opo}
P_{s+i}=(\frac{2Q_s}{Q_{s,1}}+\frac{2Q_i}{Q_{i,1}}) \frac{Q_p}{Q_{p,1}} P_{th}(\sqrt{P_p/P_{th}}-1).
\end{equation}
The theoretical predictions (solid black lines) based on Eq.\,(\ref{eq_p_opo}) are plotted in Figs.\,\ref{opo}(a) and \ref{opo}(b) and match well with the experimental data (blue circles). With an elevated on-chip near-visible pump power of 93.0\,$\mu$W, we record an off-chip infrared power of 1.5\,$\mu$W, corresponding to an on-chip OPO power of 10.5\,$\mu$W, which further translates to an absolute power conversion efficiency $\eta$, defined as $\eta=P_{s+i}/P_p$, of 11\% [Fig.\,\ref{opo}(b)]. According to Eq.\,(\ref{eq_p_opo}), the conversion efficiency reaches its maximum on condition that $P_p=4P_{th}$, where $\eta_\mathrm{max}=({2Q_s}/{Q_{s,1}}+{2Q_i}/{Q_{i,1}}) ({Q_p}/{Q_{p,1}})/4$. The efficiency depletion at the high-power regime is mainly attributed to the back-conversion of the signal and idler fields. Different coupling conditions could be adapted by varying the gap between the bus waveguide and microring for various applications. For example, a $\eta_\mathrm{max}$ of 0.25 is derived with the assumption that all three interacting modes are critical-coupled whilst strong over-coupling condition provides a $\eta_\mathrm{max}$ approaching unity. In our case, the under-coupling condition provides an ultralow threshold of 30\,$\mu$W though limits our current $\eta_\mathrm{max}$ to 11\%. 

We further demonstrate the feasibility of our ultralow-threshold OPO in wavelength tuning over a large spectrum range, which promises flexible coherent optical sources for future on-chip applications. It is well-known that, in the microring resonators, the phase-matching condition could be satisfied simultaneously by many signal-idler mode pairs, with their wavelength difference increasing at intervals of 2 times of free spectral range (FSR) for a given pump mode and temperature. However, due to the inherent dispersion, the frequency mismatch $\delta$ varies for different mode pairs, and only one pair of modes oscillates with the lowest threshold. Therefore, we investigate the tuning of signal and idler wavelengths by varying the resonator temperature as well as the pump mode number. 

\begin{figure*}[t]
\includegraphics[width=0.85\textwidth]{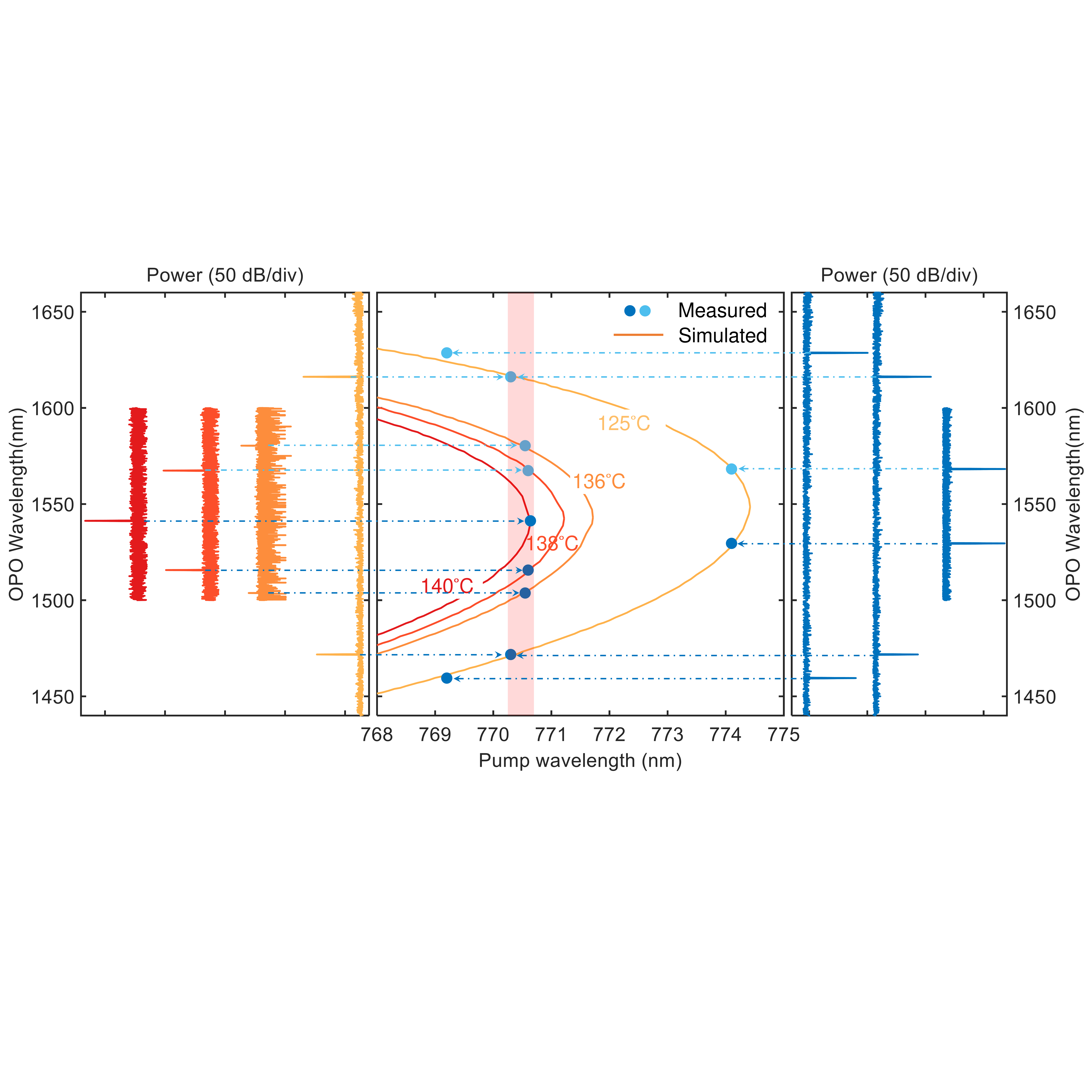}  
\caption{\label{tunability}\textbf{OPO wavelength tuning by varying both the temperature and pump wavelength.} Simulated (solid colored lines) and measured (colored circles) OPO wavelengths as a function of pump wavelength at various temperatures. As an example, the right panel plots the recorded spectra with different pump wavelengths at 125$^\circ$C. The left panel indicates the occurrence of degenerate parametric oscillation by varying the temperature.}
\end{figure*}

As indicated by the solid color lines in the middle panel of Fig.\,\ref{tunability}, the theory predicts that the quasi-phase-matched signal and idler wavelengths could be continuously tuned as a function of the pump wavelength and the chip temperature (see Methods for simulation details). However, due to the constrain of resonance condition, we can only experimentally access discrete points on the lines, as marked by the blue dots. On one hand, for a given temperature, the spectral separation between the signal and idler increases with the decreasing pump wavelengths. For example, by fixing the operation temperature at 125$^\circ$C, the OPO wavelength could be tuned from $1450$ to $1650\,\mathrm{nm}$, with the pump wavelength change of only $5\,\mathrm{nm}$. The detailed spectra are plotted on the right panel, whose signal and idler wavelengths (blue circles) are mapped to the left panel and show good agreement with the theoretical predication. On the other hand, for a given pump mode, the temperature tuning allows for the fine-control of the OPO wavelengths. As indicated on the left panel, for a fixed pump mode around $\sim$770.5\,nm (red shaded region), the signal-idler separation is observed to narrow as the temperature increases and an degenerate OPO is experimentally demonstrated at an optimal temperature of 140$^\circ$C, which match well with the prediction. Hence, our OPO device has shown good potential to realize the relatively precise control of oscillation wavelengths by combining pump wavelength and temperature tunings. Noting that the demonstrated wavelength tuning is broad but inherently discrete, continuous tuning (mode-hop free) is always demanding and remains to be investigated via several promising strategies: One approach is the material refractive index change via either thermo-optic or electro-optic effect, which allows tens of GHz tuning range~\cite{Hu:20,surya2020stable}. Another is cascading multiple microrings with varying geometry parameters on each individual bus waveguide~\cite{xiangcascaded,Surya:18}, so that the tuning range of multiple resonators could possibly overlap and cover one single FSR.

Finally, we compare the performances of various integrated platforms, including both $\chi^{(3)}$ and $\chi^{(2)}$ mechanisms, for their respective state-of-the-art cavity-enhanced OPOs in terms of the material, refractive index $n_0$, nonlinear coefficients, device structure, operating wavelengths, Q factor, and threshold power, as presented in Tab.\,\ref{tab:comparison}. 
\begin{table*}[t]
\caption{\label{tab:comparison}Performances of various nonlinear integrated platforms for microcavity-enhanced parametric oscillation.}
\begin{ruledtabular}
\begin{tabular}{cccccccccc}
 Materials& $n_0$ &$\chi^{(3)}$ ($10^{-18}\mathrm{m}^2/\mathrm{W}$)&$\chi^{(2)}$(pm/V)& Structure &Process & $\lambda_p-\lambda_s$(nm) & Q($\times 10^6$) & $P_\mathrm{th}$(mW) & Ref.\\ \hline
\multirow{2}{*}{LN} & \multirow{2}{*}{2.2} & \multirow{2}{*}{0.18} & \multirow{2}{*}{40} & \multirow{2}{*}{microring }& $\chi^{(2)}$ & 775--1550 & 0.6 & \textbf{0.030} & this work \\
& &  & & & $\chi^{(3)}$ & 1560--1540 & 1.1 & 80 & \cite{wang2019monolithic} \\
\hline
\multirow{2}{*}{AlN} & \multirow{2}{*}{2.1} & \multirow{2}{*}{0.23} & \multirow{2}{*}{6} & \multirow{2}{*}{microring } & $\chi^{(2)}$ & 780--1560 & 1.0 & 12 & \cite{Bruch_OPO} \\
& &  & & &$\chi^{(3)}$ & 1560--1530 & 1.6 & 25 & \cite{liu2018integrated} \\
\hline
\multirow{2}{*}{Si$_3$N$_4$} & \multirow{2}{*}{2} & \multirow{2}{*}{0.25} & \multirow{2}{*}{---} & \multirow{2}{*}{microring } & \multirow{2}{*}{$\chi^{(3)}$} & 1560-1540 & 37 & 0.33 & \cite{ji2017ultra} \\
& &  & & & & 900--700 & 2.5 & 0.90 & \cite{XiyuanLu_OPO} \\
\hline
AlGaAs & 3.4 & 26 & 120 & microring &$\chi^{(3)}$ & 1540--1520 & 1.5 & 0.036 & \cite{chang2020ultra} \\
\hline
GaP & 3.1 & 11 & 82 &microring & $\chi^{(3)}$ & 1560--1540 & 0.2 & 3.0 & \cite{wilson2020integrated} \\
\hline
SiC & 2.7 & 1 & 12 &microring & $\chi^{(3)}$ & 1560--1530 & 1.1 & 8.5 & \cite{SiC_opo} \\
\hline
Si & 3.5 & 10 & --- & microring &$\chi^{(3)}$ & 2600--2500 & 0.6 & 3.1 & \cite{griffith2015silicon} \\
\hline
\multirow{2}{*}{SiO$_2$} & \multirow{2}{*}{1.4} & \multirow{2}{*}{0.022} & \multirow{2}{*}{---} & microtoroid & \multirow{2}{*}{$\chi^{(3)}$} & 1550--1500 & 100 & 0.050 & \cite{del2007optical} \\
& &  & & microdisk & & 1550--1549 & 1100 & 0.95 & \cite{wu2020greater}\\
\end{tabular}
\end{ruledtabular}
\end{table*}
Theoretically, $P_{th} \propto 1/Q^2$ for the $\chi^{(3)}$ OPO~\cite{ji2017ultra}. However, in the case of $\chi^{(2)}$ process, $P_{th} \propto 1/Q^3$~[Eq.\,(\ref{eq_threshold})]. Therefore, to reach the same threshold, $\chi^{(2)}$ OPO has a less stringent requirement on its Q factor. Despite the fact that silicon photonics, including Si, Si$_3$N$_4$ and SiO$_2$, have much higher Q factors due to their mature fabrication techniques, the demonstration in this work, to the best of our knowledge, delivers the lowest threshold power among all on-chip OPOs so far. Moreover, due to the three-wave mixing nature, $\chi^{(2)}$ OPO offers broader spectral-separation between the pump and signal(idler) fields intrinsically, which promises for sufficient on-chip filtering. We note that piezoelectric III–V materials, such as Al(Ga)N, (Al)GaAs and GaP, feature both significant $\chi^{(2)}$ and $\chi^{(3)}$ nonlinear coefficients, while relying on challenging waveguide geometrical engineering to achieve modal phase matching between fundamental and high-order modes which limits their OPO performance as well as operation wavelengths. Overall, the periodically poled thin-film lithium niobate microresonator prove to be an efficient OPO with the lowest power threshold. In addition, due to its flexibility in ferroelectric domain control for quasi-phase matching and greatly enhanced light-matter interaction afforded by its high effective $\chi^{(2)}$ nonlinearity and Q factor, our device on a integrated photonic chip opens new possibilities for various applications and alternative device designs could be envisioned based on target applications. Firstly, PPLNMR device promises chip-based tunable classical and quantum light sources. For example, the poling grating number $M$ could be tailored to realize parametric oscillation at the mid-infrared wavelengths for the molecule spectroscopy~\cite{tittel2003mid}. Over-coupling the degenerate signal modes could lead to efficient on-chip squeezers, which play a crucial role in continuous variable quantum information processing~\cite{PhysRevLett.124.193601}. Secondly, due to the multiple modes and nonlinear optical processes present in the same device, the OPO dynamics could be an indicator for novel physics of light-matter interactions in the microresonator~\cite{mingenhance}. For example, inspired by the recent demonstration on AlN~\cite{Bruch2020}, ultra-efficient pockels soliton microcomb could be foreseen in the same device with the efforts of coordinating its Kerr, quadratic, and photorefractive effects~\cite{he2019self}. Lastly, by coupling several PPLNMRs, significant nonlinear interplay between the OPOs offers an platform for realizing the photonic neural networks and Ising models, which find applications in artificial machine learning~\cite{pernice_neuronetwork,doi:10.1063/5.0016140}.  

In conclusion, we demonstrate $\chi^{(2)}$ optical parametric oscillation in lithium niobate nanophotonics. Leveraging the high optical confinement, as well as optimized quasi-phase matching condition, an ultra-low power threshold (30\,$\mu$W) of optical parametric oscillation is obtained and an on-chip conversion efficiency of 11\,\% is measured with an on-chip pump power of 93\,$\mu$W. The tunability of the OPO wavelengths is also investigated via the pump-wavelength and temperature tuning, and the experimental results indicate a wide tuning range. By further pushing the Q factor to that of bulk LN WGRs, approaching $10^8$, a pump threshold down to several picowatts could be attainable in PPLNMR devices, allowing the study of OPO at a few-photon level and is critical for the scalability of photonic quantum systems. In light of the most recent works on near-octave comb generation, efficient SHG, and bright photon pair generation, PPLNMR devices hold promise for the chip-based tunable coherent light sources as well as the monolithic realization of nonlinear and quantum photonics.

\section*{Methods}
\noindent \textbf{Nanofabrication.} 
The devices are fabricated on a commercial LNOI wafer (supplied by NANOLN) with 610\,nm thick Z-cut LN thin film on 1.8\,$\mu$m silicon dioxide (SiO$_2$) on a silicon substrate. The bus waveguide is ultimately tapered to a width of 4\,$\mu$m at both facets to improve the fiber-to-chip coupling efficiency. The pattern is defined by a 100\,kV electron beam lithography system (Raith EBPG 5000+) with a negative FOx-16 resist, which is then developed in 25$\%$ TMAH solution to ensure a high contrast. The exposed pattern is transferred onto the LN thin film using an optimized inductively couple plasma (ICP) reactive ion etching (RIE) process with Ar$^+$ plasma. For the subsequent poling process, the radial nickel electrodes are initially patterned on top of the LN microring via the lift-off process. The periodic domain inversion is then enabled by keeping the silicon substrate as the electrical ground while applying several 600\,V, 250\,ms pulses on the electrodes at an elevated temperature of 250\,$^\circ$C. After removing the nickel electrodes, the chip is cleaved to expose the waveguide facets for fiber-to-chip coupling. The insertion losses are calibrated to be 8.4 and 11.1\,dB/facet for the infrared and near-visible lights, respectively.

\vspace{1 mm}
\noindent \textbf{Numerical simulation.} 
The parametric signal and idler wavelengths as a function of the pump wavelength and temperature are numerically investigated using a commercial finite-difference-eigenmode solver (Lumerical MODE) based on the equations:
\begin{gather}
\frac{1}{\lambda_p} = \frac{1}{\lambda_s}+\frac{1}{\lambda_i}, \label{energy}\\
\frac{n_p(T)}{\lambda_p} = \frac{n_s(T)}{\lambda_s}+\frac{n_i(T)}{\lambda_i}+\frac{M}{2\pi R}, \label{momentum}
\end{gather}
where $\lambda_{p/s/i}$ and $n_{p/s/i}$ denote the wavelengths and refractive indices for the pump, signal, and idler lights, respectively. In the simulation, we used the Sellmeier equation for congruent LN in Ref.~\cite{Zelmon:97} and its wavelength-dependent thermo-optic coefficient is also considered~\cite{doi:10.1063/1.1988987}.

\def\bibsection{\section*{\textbf{references}}}
\bibliographystyle{myaipnum4-1}
\bibliography{References}% Produces the bibliography via BibTeX.
\vspace{1 mm}
\noindent \textbf{Acknowledgements.} This work is supported by Department of Energy, Office of Basic Energy Sciences, Division of Materials Sciences and Engineering under Grant DE-SC0019406. The authors thank Michael Rooks, Yong Sun, Sean Rinehart, and Kelly Woods for assistance in device fabrication.

% \vspace{1 mm}
% \noindent \textbf{Author contributions.} J.L. performed the device design, fabrication and measurement with the assistance from A.S., Z.G., J.S., C.-L.Z.. J.L. and H.X.T. prepared the manuscript in discussion with all authors. H.X.T supervised the project.

\vspace{1 mm}
\noindent \textbf{Competing interests.} The authors declare no competing interests.

\end{document}